# Design of Rate-Compatible Serially Concatenated Convolutional Codes

Alexandre Graell i Amat[1,2], Fredrik Brännström[3], and Lars K. Rasmussen[4]

[1]GET/ENST Bretagne, Electronics Department, CS 83818 - 29238 Brest Cedex 3, France, alexandre.graell@enst-bretagne.fr
[2]Department of Technology, Universitat Pompeu Fabra, Passeig Circumval.lació 8, 08003 Barcelona, Spain, alex.graell@upf.edu
[3]Department of Signals and Systems, Chalmers University of Technology, SE-412 96 Göteborg, Sweden, fredrikb@chalmers.se
[4]ITR, University of South Australia, Mawson Lakes, SA 5095, Australia, lars.rasmussen@unisa.edu.au

## Abstract

Recently a powerful class of rate-compatible serially concatenated convolutional codes (SCCCs) have been proposed based on minimizing analytical upper bounds on the error probability in the error floor region. Here this class of codes is further investigated by combining analytical upper bounds with extrinsic information transfer charts analysis. Following this approach, we construct a family of rate-compatible SCCCs with good performance in both the error floor and the waterfall regions over a broad range of code rates.

## 1 Introduction

With the invention of Turbo Codes [1] and the emergence of turbo-like codes in general, channel coding has finally closed the gap to the capacity for an additive white Gaussian noise (AWGN) channel. We are now able to approach the AWGN capacity to within a few fractions of a decibel [2, 3].

Although the capacity-approaching performance of turbo-like codes has been a major step forward, there is still a practical need for improvements in terms of versatility, throughput and simplicity. Techniques such as adaptive modulation and automatic repeat-request (ARQ) coding schemes are becoming important parts of modern wireless communication systems. These techniques require versatile rate-compatible code schemes in terms of both code rate and block length. Practical limitations further require low complexity and minimal performance losses as compared to dedicated fixed rate codes.

Puncturing is an effective strategy for designing rate-compatible code families such as rate-compatible parallel concatenated convolutional codes, serially concatenated convolutional codes (SCCCs) and low density parity check code structures, respectively [4, 5]. The code rate flexibility provided by rate-compatible codes is, however, generally obtained at the expense of performance losses for the higher rate schemes. To compensate for these losses, a more powerful mother code is required, which in turn leads to more complex designs.

A powerful class of rate-compatible SCCCs with low complexity has been proposed recently in [6, 7]. In contrast to classical SCCCs, characterized by the concatenation of an outer code with a rate $R_\mathrm{I} \leq 1$ inner code, the codes proposed in [6, 7] allow the inner code to be punctured beyond the unitary rate. These codes offer good performance over a wide range of code rates, using simple constituent codes. The key idea is to replace puncturing from the outer code with puncturing of the inner code, maintaining a constant block length of the outer code and its distance spectrum properties and thus keeping a constant interleaver gain for all code rates. Significantly lower error floors are obtained even for high rate codes.

Design criteria for this new class of SCCCs were suggested in [6, 7] based on union bounds (UBs) for the error probability [8]. The code optimization criteria proposed in [6] target the error floor (EF) region, leading to correspondingly good code constructions for this region. This approach, however, provides little insight into the code convergence behavior in the waterfall (WF) region.

In this work, we analyze the convergence properties of the SCCC proposed in [6, 7] using extrinsic information transfer (EXIT) chart techniques. Based on the EXIT chart analysis, design criteria are suggested for optimizing the performance in the WF region. In most cases, the optimal design criteria for the WF and EF regions do not coincide. In fact for low code rates optimizing the performance in the WF region leads to the worst possible performance in the EF region and vice versa. In order to provide a suitable trade-off, a joint approach is required. Combining EXIT charts with UBs techniques we construct a family of low-complexity rate-compatible SCCCs with good performance in both EF and WF regions.

A. Graell i Amat is supported by the Spanish Ministry of Education and Science under JdC grant and by the Australian Research Council (ARC) Communications Research Network (ACoRN) under ARC grant RN0459498. A. Graell i Amat and F. Brännström are supported in part by the EU 6th FP Network of Excellence in Communications (NEWCOM), contract no. 507325. L. K. Rasmussen is supported by the Australian Government under ARC Grant DP0558861. F. Brännström and L. K. Rasmussen are also supported by the Swedish Research Council under Grant 723-2002-4533.

## 2 System Model

A block diagram of the SCCC from [6, 7] is shown in Fig. 1. The binary information data is collected in a vector $\mathbf{u} \in \{+1, -1\}^K$. The data is first encoded by an encoder $\mathcal{C}_a$, punctured by a fixed puncturer $\mathcal{P}_a$ and interleaved before being forwarded to encoder $\mathcal{C}_b$. A rate-1/2, 4-state, systematic convolutional code (CC), denoted by CC(1, 5/7) in octal notation, is chosen for both $\mathcal{C}_a$ and $\mathcal{C}_b$. The output of $\mathcal{C}_b$ is then punctured by the puncturers $\mathcal{P}_b^s$ and $\mathcal{P}_b^p$ for systematic bits and parity bits, respectively. Encoder $\mathcal{C}_a$ together with puncturer $\mathcal{P}_a$ are referred to as the *outer code* $\mathcal{C}_O$ and encoder $\mathcal{C}_b$ together with puncturers $\mathcal{P}_b^s$ and $\mathcal{P}_b^p$ are referred to as the *inner code* $\mathcal{C}_I$. MUX represents a multiplexer, converting two (or more) streams of bits into a singe bit stream.

A puncturer is defined by a puncturing pattern with a certain pattern length $N_p$. For example, the fixed outer puncturer, $\mathcal{P}_a$, is chosen to puncture every other parity bit. Using standard notation, the corresponding pattern is described by $\begin{bmatrix} 1 & 1 \\ 1 & 0 \end{bmatrix}$ and has $N_p = 4$. For any codeword of $\mathcal{C}_a$ of length $N_a$, this pattern is then repeated $N_a/N_p$ times.

In this paper, we refer to puncturing patterns as vectors, rather than matrices. The pattern above is therefore referred to as $\mathbf{p} = \begin{bmatrix} p_0 & p_1 & p_2 & p_3 \end{bmatrix} = \begin{bmatrix} 1 & 1 & 1 & 0 \end{bmatrix}$. Alternatively, the pattern is described by a sequence listing the indices of the zero entries in the pattern vector, i.e., in the case above this sequence is $\{3\}$. Using this notation, a rate-compatible code family can be described by a table of indices, representing the order in which bits in a codeword are to be punctured.

The SCCC in [6, 7] uses the fixed outer puncturer detailed above. Since the rate of the outer code $\mathcal{C}_O$ is then 2/3, the interleaver length is $N = 3K/2$, where $K$ is the information block length.

An equivalent block diagram of the SCCC in Fig. 1 is shown in Fig. 2. As explained later on, the description in Fig. 2 allows us to decouple the effects of systematic bits, outer parity bits and inner parity bits on the performance. A similar description of serially concatenated codes was used in [9] and [10] for convergence analysis and in [6, 7] for deriving upper bounds on the error probability.

In Fig. 2, $\mathcal{C}_1$ and $\mathcal{C}_2$ are now the rate-1 CC(5/7), corresponding to the parity part of $\mathcal{C}_a$ and $\mathcal{C}_b$ from Fig. 1, respectively. The fixed puncturer $\mathcal{P}$ in Fig. 2 has the pattern $\mathbf{p} = \begin{bmatrix} 1 & 0 \end{bmatrix}$. Furthermore, interleaving the combination of $\mathcal{P}_0$ and $\mathcal{P}_1$ in Fig. 2 translates into the equivalent $\mathcal{P}_b^s$ in Fig. 1, while $\mathcal{P}_2$ is identical to $\mathcal{P}_b^p$. The systematic bits of $\mathcal{C}_I$ in Fig. 1 contain both $\mathbf{x}_0$ and $\mathbf{x}_1$ from Fig. 2, whereas $\mathbf{x}^p$ is equal to $\mathbf{x}_2$. Note also that the vectors $\mathbf{v} \in \{+1, -1\}^N$ and $\mathbf{z} \in \{+1, -1\}^N$ in Fig. 1 and Fig. 2 contain the same bits. Furthermore, in Fig. 2 the two dashed boxes define two corresponding codes referred to as the *upper code* $\mathcal{C}_U$ and the *lower code* $\mathcal{C}_L$, respectively.

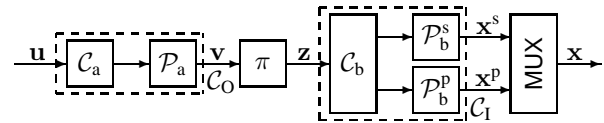

Fig. 1. Classical block diagram of a serially concatenated code.

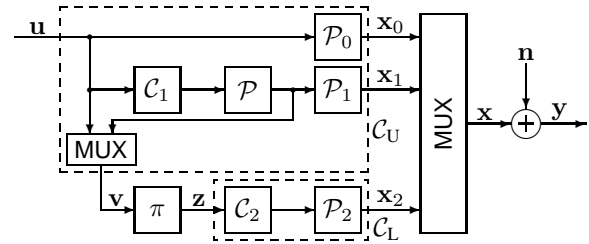

Fig. 2. Equivalent block diagram of a serially concatenated code.

The permeability rate is defined as the fraction of bits remaining after puncturing. In Fig. 2 $\rho_0$, $\rho_1$, and $\rho_2$ denote the permeability rates for the upper systematic bits, the upper parity bits, and the lower parity bits, respectively. Finally, the systematic information bits $\mathbf{x}_0 \in \{+1, -1\}^{K\rho_0}$ are multiplexed with the parity bits $\mathbf{x}_1 \in \{+1, -1\}^{K\rho_1/2}$ and $\mathbf{x}_2 \in \{+1, -1\}^{3K\rho_2/2}$ to form the transmitted codeword $\mathbf{x} = [\mathbf{x}_0, \mathbf{x}_1, \mathbf{x}_2] \in \{+1, -1\}^L$. Here $L = K/R$ is the total number of transmitted bits and $R$ is the rate of the overall code given by

$$R = \frac{K}{L} = \left(\rho_0 + \frac{1}{2}\rho_1 + \frac{3}{2}\rho_2\right)^{-1}. \quad (1)$$

The received matched filter output $\mathbf{y}$ shown in Fig. 2 is $\mathbf{y} = \mathbf{x} + \mathbf{n}$, where each element in $\mathbf{n}$ is zero-mean Gaussian with variance $\sigma^2 = \frac{N_0}{2RE_b}$.

## 3 Code Design Criteria

The error probability performance of turbo-like codes is characterized by a steep transition region, the so-called *waterfall* region, followed by a more gently sloping plateau region, referred to as the *error floor* region. Two different techniques have been suggested for designing turbo-like codes with good performance in the two different regions. One approach relies on analytical bounding techniques based on the assumption of uniform interleavers [8], and leads to design rules for the constituent encoders. As the bounds provide accurate predictions of the EF, this technique is aimed at improving code performance in the EF region. A different strategy based on EXIT charts or density evolution [11, 12] is aiming at improving the code performance in the WF region. These techniques predict the convergence behavior of the iterative decoding process based on the separate simulation of the constituent codes assuming infinite length interleavers. The EXIT charts approach can be used in a search for codes exhibiting good behavior in the WF region.

The goal in the design of turbo-like schemes is to find codes that perform well in both the EF region and the WF region. This suggests that a joint design approach may be appropriate. Unfortunately, design criteria derived from analytical bounds and from EXIT charts, respectively, are in general competing rather than complementing, i.e., improving code performance in the EF region leads to a penalty in terms of convergence threshold, and vice versa.

For the class of SCCCs of Fig. 2 the error performance depends on the puncturing patterns of $\mathcal{P}_0$, $\mathcal{P}_1$ and $\mathcal{P}_2$ and, subsequently, on the permeability rates $\rho_0$, $\rho_1$ and $\rho_2$. Therefore, the objective is to optimize these parameters to ensure good performance in the EF and WF regions. In [6, 7] bounds on the error probability for this class of SCCCs were derived. Based on the bounds, design criteria were proposed for optimizing the puncturing patterns in $\mathcal{P}_b^s$ and $\mathcal{P}_b^p$ in terms of minimizing the probability of error in the EF region. Following these criteria a family of rate-compatible SCCCs was proposed.

A strategy similar to the one proposed in [6, 7] was proposed in [10] for constructing rate-compatible multiple concatenated block codes. Here, the optimization of the puncturing patterns and the permeability rates is based on both upper bounds assuming uniform interleavers, and on EXIT functions assuming infinite interleavers. The design procedure based on EXIT functions is also suggested in [9, 13] for multiple concatenated codes. The permeability rates are there optimized in terms of minimizing the convergence threshold in the WF region, i.e., minimizing the required SNR to reach a specific target BER. However, the optimized permeability rates found in [9, 13] assume random puncturing patterns.

Here, instead of assuming random puncturing patterns, we evaluate the EXIT functions for the component codes using the puncturing patterns found in [6]. Note that finding the puncturing patterns in $\mathcal{P}_0$, $\mathcal{P}_1$, and $\mathcal{P}_2$ for minimizing the convergence threshold using EXIT charts is prohibitively complex. For these types of SCCCs, even the sub-optimal incremental approach used in [6] is exceedingly complex for optimization of puncturing patterns based only on EXIT charts. The key idea in this paper is therefore to keep the rate-compatible puncturing patterns found in [6] for minimizing the EF. Based on these patterns, EXIT charts are then applied to find the permeability rates $\rho_0$, $\rho_1$, and $\rho_2$ minimizing the convergence thresholds. In this way, we obtain a family of rate-compatible SCCCs that offer good performance in both the EF and the WF regions.

## 4 Convergence Analysis

Standard EXIT charts for SCCCs [11] analyze the convergence properties of the code by plotting the EXIT curves for the outer and inner codes, respectively, ($\mathcal{C}_O$ and $\mathcal{C}_I$ in Fig. 1) in a single diagram. Whereas this approach works well for standard SCCCs, it fails when the inner code in Fig. 1 is systematic. This behavior is explained by two observations. Firstly, EXIT charts based on the outer and inner code in Fig. 1 treat the inner encoder with its puncturers as a unique entity separated from the outer encoder, thus diluting the effect of systematic bits and parity bits on the performance. Secondly, the dependency of the puncturer $\mathcal{P}_b^s$ on the outer decoder, which has a crucial impact on the performance (see [6]), is lost. Ignoring these dependencies lead to pessimistic predictions of convergence thresholds, which can significantly diverge from the actual behavior of the decoding process.

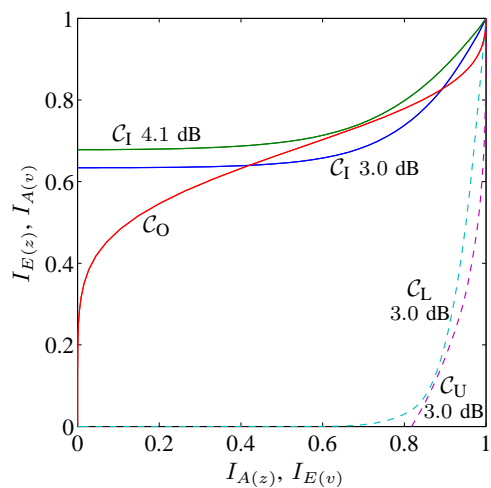

Fig. 3. EXIT charts based on the schemes of Fig. 1 (solid curves) and Fig. 2 (dashed curves) for a rate-5/6 code with $\rho_0 = 200/200$, $\rho_1 = 20/100$, and $\rho_2 = 20/300$.

As it was done in [7] for the derivation of the error bounds, this problem can be solved by analyzing the convergence behavior of the SCCC based on the equivalent scheme of Fig. 2. The convergence analysis can therefore be obtained by plotting the EXIT curves for $\mathcal{C}_U$ and $\mathcal{C}_L$ in the same diagram, as suggested in [9, 10]. Note that the two curves in the EXIT chart for $\mathcal{C}_U$ and $\mathcal{C}_L$ both depend on the signal-to-noise ratio (SNR), in contrast to the EXIT chart for $\mathcal{C}_O$ and $\mathcal{C}_I$ where only the curve for $\mathcal{C}_I$ depends on SNR.

To exemplify this behavior, we plot in Fig. 3 the EXIT charts for a rate-5/6 SCCC, using the puncturing patterns reported in [6] with permeability rates $\rho_0 = 200/200$, $\rho_1 = 20/100$ and $\rho_2 = 20/300$. $I_{A(v)}$ denotes the prior mutual information for both $\mathcal{C}_O$ and $\mathcal{C}_U$, while $I_{E(v)}$ denotes the extrinsic mutual information for the same components. In a similar way, $I_{A(z)}$ and $I_{E(z)}$ denote the prior and extrinsic mutual information for both $\mathcal{C}_I$ and $\mathcal{C}_L$.

The solid curves in Fig. 3 correspond to the evolution of the mutual information for the outer and inner codes in Fig. 1, i.e., standard EXIT charts. For $E_b/N_0 = 3.0$ dB the trajectories of the two curves intersect, preventing good performance at this $E_b/N_0$. The tunnel between the two curves opens at $E_b/N_0 = 4.1$ dB, indicating a convergence threshold around this value. The convergence threshold predicted by this EXIT chart however, does not agree with the actual behavior of this code. In fact, for the simulated performance, the decoding converges at significantly lower SNR (around $E_b/N_0 = 3.0$ dB). The EXIT charts for the same rate-5/6 code based on the scheme of Fig. 2 is also shown in Fig. 3. The dashed curves, plotted at $E_b/N_0 = 3.0$ dB, show a tunnel, allowing the convergence of the iterative decoding. The predicted convergence threshold is in perfect agreement with the simulated performance.

# 5 Rate-Compatible SCCC

If the upper code has a code rate higher than one, the invertibility of the overall code cannot always be assured. Therefore the code rate of the upper code is restricted to be lower or equal to one, i.e., the total number of bits in $\mathbf{x}_0$ and $\mathbf{x}_1$ must be greater than or equal to $K$. $\mathcal{P}_0$ and $\mathcal{P}_1$ for the upper code can be optimized jointly. However, restricting the upper code to transmit all systematic bits leads, for this setup, to better or similar results than if systematic bits are allowed to be punctured. Therefore, hereafter $\rho_0$ is fixed to one.

As in [6, 7] we assume a block length of $K = 200$ information bits, corresponding to an interleaver length of $N = 300$ bits. The length of the puncturing patterns in $\mathcal{P}_0$, $\mathcal{P}_1$, and $\mathcal{P}_2$ are fixed accordingly to 200, 100, and 300, respectively. For larger block lengths, the puncturing patterns are just repeated. To simplify notation we introduce a vector $\mathbf{D} = [d_0, d_1, d_2]$ containing the number of transmitted bits in $\mathbf{x}_0$, $\mathbf{x}_1$, and $\mathbf{x}_2$, respectively, i.e., $d_0 = 200\rho_0$, $d_1 = 100\rho_1$, $d_2 = 300\rho_2$, and $L = d_0 + d_1 + d_2 = 200/R$. Note that if $d_2$ is chosen for a specific code rate $R = 200/L$, $\mathbf{D}$ is specified since $d_0 = 200$ and $d_1 = L - 200 - d_2$.

The puncturing patterns for $\mathcal{C}_U$ and $\mathcal{C}_L$ are optimized separately in [7]. For each code rate, $\mathcal{P}_2$ is selected to optimize the *input-output weight enumerating function* (IOWEF) of $\mathcal{C}_L$, while satisfying the rate-compatibility restriction. To this purpose, a suboptimal algorithm that works incrementally, puncturing one bit at a time, is used. The best puncturing pattern for $\mathcal{C}_L$, allowing puncturing of only one parity bit is found by comparing the IOWEFs of all 300 possible puncturing patterns [6, 7]. Since the code should be rate-compatible, all bits that are kept at a certain rate $R$ must include all bits used for all rates higher than $R$. Applying this restriction the best pattern puncturing two parity bits is found by comparing the IOWEFs of all 299 possible puncturing pattern where the first punctured bit is fixed from the previous search. This sub-optimal method of finding good rate-compatible puncturing patterns for the performance in the EF region can be continued until the order of all 300 possible bits to be punctured is found. The result is a table of 300 indices stating the order the parity bits of the lower code should be punctured. The best puncturing for $\mathcal{C}_U$, is found in a similar way by optimizing the *output weight enumerating function* of $\mathcal{C}_U$. The tables of indices for the upper code (100 indices) and the lower code (300 indices) are reported in [6]. These tables only specify the order to puncture the bits for the upper and lower code, but not the permeability rates. For a given code rate, there are several choices of $\rho_1$ and $\rho_2$. For example, for $R = 2/3$ the total number of bits transmitted should be $L = 300$ and $d_2$ can be chosen between $0 \leq d_2 \leq 100$ ($0 \leq \rho_2 \leq 100/300$). The results in [6, 7] show that to minimize the EF, $\rho_2$ should be kept as high as possible for all given code rates, similar to the conclusions drawn in [10] for serially concatenated single parity check codes.

As mentioned earlier on, the optimization of the puncturing patterns in $\mathcal{P}_1$ and $\mathcal{P}_2$ based on EXIT functions is exceedingly complex. Instead, we keep the puncturing patterns found in [6] to ensure good performance in the EF, and then apply EXIT functions to find the permeability rates, $\rho_1$ and $\rho_2$, minimizing the convergence thresholds. An EXIT function of a component code contains all necessary information to perform the convergence analysis for all SNRs and code rates [9, 14]. The projection of two (or more) of these EXIT functions onto a two-dimensional EXIT chart predicts the required SNR to reach a target BER [9, 14]. To this end 101 EXIT functions are calculated for the upper code, one for each possible $d_1 = 0, 1, 2, \ldots, 100$. For the lower code 301 EXIT functions are determined, one for each $d_2 = 0, 1, 2, \ldots, 300$. Projecting all combinations of the 101 upper EXIT functions with the 301 lower EXIT functions predicts the required SNR to reach a specified target BER (here chosen to be $P_b = 10^{-5}$) for all code rates $1/3 \leq R \leq 1$ of the SCCC in Fig. 2 [9, 14].

Fig. 4 shows the required SNR for some of these code rates. The upper set of curves shows the required SNR to reach $P_b = 10^{-9}$ in the EF region predicted by the UBs derived in [6, 7]. Here, it is clear that the minimum required SNR is achieved if $d_2$ is maximized. For $R = 1/2$, $L = 400$ and $d_2$ must be in the range $100 \leq d_2 \leq 200$, since $d_0 = 200$ and $0 \leq d_1 \leq 100$. For this code rate, $E_b/N_0 = 4.2$ dB is required to reach $P_b = 10^{-9}$ when $d_2 = 200$. The lower set of curves in Fig. 4 shows the required SNR to reach $P_b = 10^{-5}$ in the WF region predicted by the EXIT chart analysis. For $R = 1/2$ the results show that the minimum required SNR is $E_b/N_0 = 1$ dB and achieved when $d_2 = 100$. Hence, the best choice of $d_2$ in the WF region is 100 and the best choice in the EF region is 200, which are the two extreme choices for $R = 1/2$. For $R = 2/3$, the optimum $d_2$ for the WF region is 50 and the optimum $d_2$ for the EF region is 100. This suggests that $d_2$ should not be chosen below 50 when

$R = 1/2$ since that will degrade the performance in both the EF and the WF region.

The optimum values of $d_2$ for all code rates for both the EF and the WF regions are shown in Fig. 5. If $d_2$ is chosen to have a value close to the line with squares, the performance is good in the EF region ($d_2$ is maximized, i.e., $d_1$ is minimized). If $d_2$ is chosen to have a value close to the line with circles, the performance is good in the WF region ($d_2 = d_1$ as long as possible). Any non-decreasing curve between the circles and the squares give a rate-compatible code with a performance somewhere in between the two extreme cases. It is also possible to choose a value of $d_2$ between the dashed line and the line with circles, but that will give worse performance in both regions. A good and simple compromise is to choose the straight line marked with triangles. This straight line corresponds to choosing

$$d_2 = \frac{3L - 600}{4}, \qquad \rho_2 = \frac{1-R}{2R}. \qquad (2)$$

The minimum required SNR can also be plotted versus all code rates $1/3 \leq R \leq 1$ as in Fig. 6. The left set of curves in Fig. 6 corresponds to the minimum required SNR to reach $P_b = 10^{-5}$ predicted by the EXIT chart analysis, and the right set of curves corresponds to the minimum required SNR to reach $P_b = 10^{-9}$ predicted by the UB technique for the three different strategies of choosing $d_2$ reported in Fig. 5. Fig. 6 also makes it clear that it is not possible to choose $d_2$ such that the performance is optimized in both the EF and the WF regions. However, choosing $d_2$ according to (2) provides a suitable compromise.

## 6 Simulation Results

The rate-compatible SCCCs discussed in this paper offer good performance over a broad range of code rates and block lengths, with only limited complexity.

In Fig. 7 we plot the performance of the SCCC code for several code rates ranging from $1/2$ to $20/21$. The permeability rates $\rho_1$ and $\rho_2$ have been chosen to minimize the decoding convergence threshold while satisfying the rate-compatibility restriction, i.e., they correspond to the WF curve in Fig. 5. A block size $K = 2000$ bits ($N = 3000$) and random interleavers are assumed. The performance after 10 (solid curves) and 20 (dashed curves) iterations are given. In the same figure, the dashed-dotted curves are the analytical upper bounds of the error probability in the EF region combined with the BER prediction in the WF region based on EXIT charts and infinite interleavers. Even though a moderate block length is used, good performance is obtained for all rates in both the EF and WF regions. Moreover, it is observed that increasing the code rate only leads to a higher convergence threshold and not to a higher EF. Even for $R = 9/10$ the predicted EF is reached at the same $P_b \approx 10^{-7}$ as for $R = 1/2$. This *flat* behavior stems from the fact that by moving the heavy puncturing from the outer encoder to the inner encoder, the interleaver gain for low rates is also kept for high rates, leading to a low and similar floor for all code rates. Furthermore, it is worth to point out that the EF will be significantly lower if a designed interleaver is used instead of random interleavers. For an S-random interleaver the EF is still not reached at $P_b = 10^{-9}$. We stress the fact that the scheme is flexible in terms of code rate, since a change of rate just corresponds to more or less puncturing while the decoding complexity is the same for all rates.

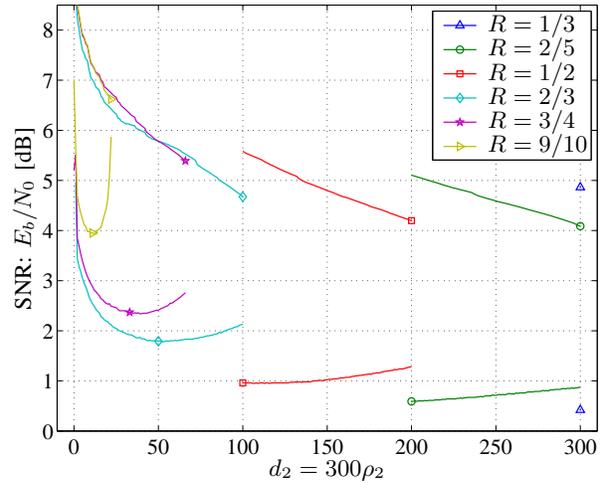

Fig. 4. Required SNR for different $\rho_2$ and code rates $R$. The lower set of curves is using EXIT charts to reach $P_b = 10^{-5}$ after 10 iterations in the WF region. The upper set of curves is using UB to reach $P_b = 10^{-9}$ in the EF region for a block length of $K = 2000$.

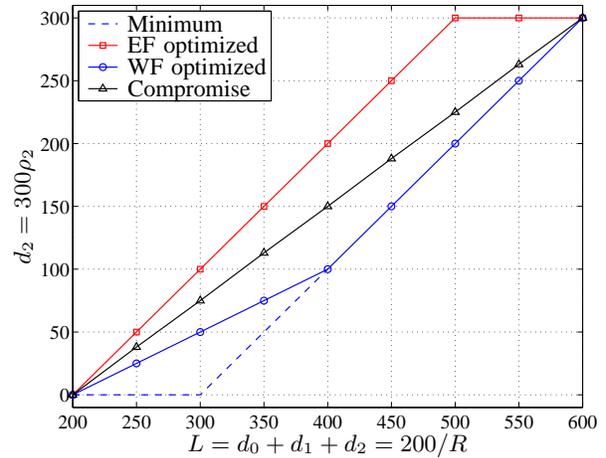

Fig. 5. Values of $\rho_2$ for different strategies and code rates $R$.

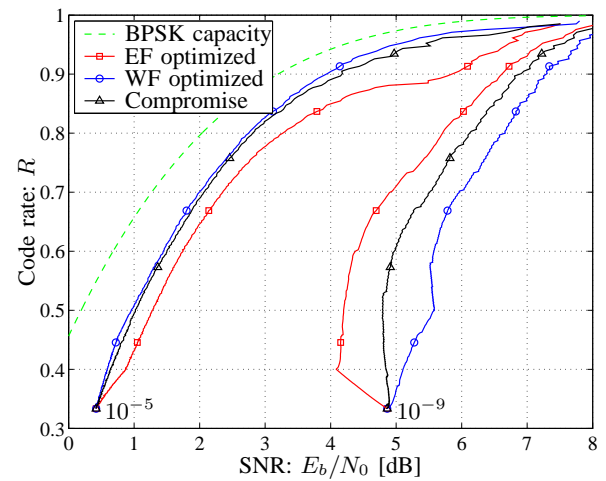

Fig. 6. Minimum required SNR for different code rates $R$. The left set of curves is using EXIT charts to reach $P_b = 10^{-5}$ after 10 iterations in the WF region. The right set of curves is using UB to reach $P_b = 10^{-9}$ in the EF region for a block length of $K = 2000$.

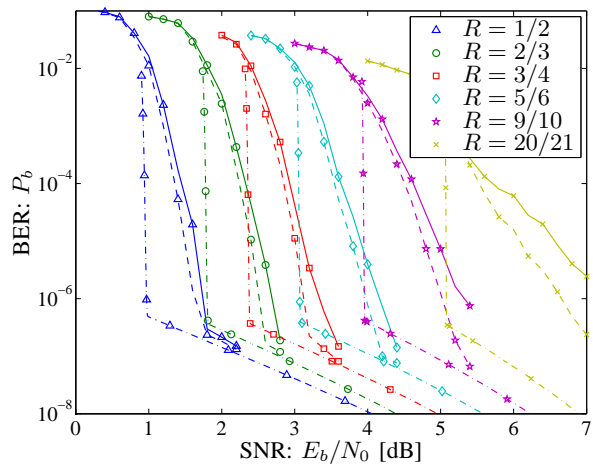
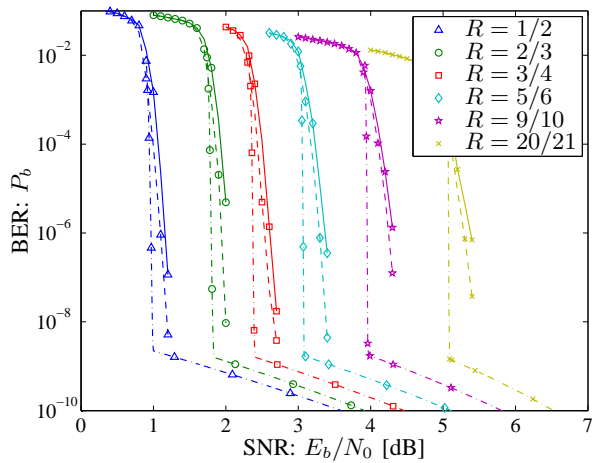

Fig. 7. Performance after 10 (solid curves) and 20 (dashed curves) iterations with $\rho_2$ optimized for the WF region. The dashed-dotted curves are the BER predictions in the WF region using EXIT charts and 10 iterations combined with the UB in the EF region for a block length of $K = 2000$ information bits.

Fig. 8. Performance after 10 (solid curves) and 20 (dashed curves) iterations with $\rho_2$ optimized for the WF region. The dashed-dotted curves are the BER predictions in the WF region using EXIT charts and 10 iterations combined with the UB in the EF region for a block length of $K = 16400$ information bits.

The performance for the same codes with $K = 16400$ bits ($N = 24600$ bits) and random interleavers are plotted in Fig. 8 together with the analytical bounds and BER predictions. For all rates, the EF region is reached at $P_b \approx 10^{-9}$. At $P_b = 10^{-5}$, all the curves for 20 iterations are around 0.9 dB from capacity.

## 7 Conclusions

In this paper, we discuss design issues of a new class of SCCCs. We analyze the convergence behavior using EXIT chart techniques and combine them with bounding techniques to construct a family of rate-compatible codes with good performance in both the waterfall and the error floor region. The proposed code structure has a low constant complexity over all code rates at the same time as it supports a broad range of code rates and block lengths. For a block length of $K = 16400$ bits the required SNR to reach $P_b = 10^{-5}$ is around 0.9 dB from the BPSK channel capacity for all rates, and the error floor with random interleavers is reached at $P_b \approx 10^{-9}$.